\documentclass[preprint,12pt]{elsarticle}

\usepackage{graphicx,ulem}
\usepackage{longtable}
\usepackage{float}
\usepackage{dcolumn}
\usepackage{graphics,epsfig}
\usepackage{amsmath,amssymb,latexsym,mathrsfs}
\usepackage{bm}
\usepackage{color}
\usepackage{color}

\journal{Physics Letters B}

\newcommand{\MeV}{\,\mathrm{MeV}}

\def\he4{$^4$He}
\def\h2{$^2$H}

\begin{document}
%\preprint{}

\begin{frontmatter}


\title{New constraints on primordial gravitational waves from Planck 2015}

\author{Luca Pagano$^{1,2}$, Laura Salvati$^2$ and Alessandro Melchiorri$^2$}
\address{$^1$ Institut d'Astrophysique Spatiale, CNRS, Univ. Paris-Sud, Universit\'{e} Paris-Saclay, B\^{a}t. 121, 91405 Orsay cedex, France}
\address{$^2$ Physics Department and INFN, Universit\`a di Roma ``La Sapienza'', Ple.\ Aldo Moro 2, 00185, Rome, Italy}

%\email{luca.pagano@roma1.infn.it}
%\affiliation{Physics Department and INFN, Universit\`a di Roma 
%	``La Sapienza'', Ple.\ Aldo Moro 2, 00185, Rome, Italy}

%\author{Laura Salvati}
%\email{laura.salvati@roma1.infn.it}	
%\affiliation{Physics Department and INFN, Universit\`a di Roma 
%	``La Sapienza'', Ple.\ Aldo Moro 2, 00185, Rome, Italy}

%\author{Alessandro Melchiorri}
%\affiliation{Physics Department and INFN, Universit\`a di Roma 
%	``La Sapienza'', Ple.\ Aldo Moro 2, 00185, Rome, Italy}

%%
%\date{\today}

\begin{abstract}
We show that the new precise measurements of Cosmic Microwave Background (CMB) temperature and polarization anisotropies made by the Planck satellite significantly improves previous constraints on the cosmic gravitational waves background
(CGWB) at frequencies $f>10^{-15}$ Hz.  On scales smaller than the horizon at the time of decoupling, primordial gravitational waves contribute to the total radiation content of the Universe. Considering adiabatic perturbations, CGWB affects temperature and polarization CMB power spectra and matter power spectrum in a manner identical to relativistic particles. Considering the latest Planck results we constrain the CGWB energy density to $\Omega_{\rm gw} h^2 <1.7\times 10^{-6} $ at 95\% CL. Combining CMB power spectra with lensing, BAO and primordial Deuterium abundance observations, we obtain $\Omega_{\rm gw} h^2 <1.2\times 10^{-6} $ at 95\% CL, improving previous Planck bounds by a factor 3 and the recent direct upper limit from the LIGO and VIRGO experiments a factor 2. A combined analysis of future satellite missions as COrE and EUCLID could improve current bound
by more than an order of magnitude.
%by a factor $\sim15$.
\end{abstract}

\end{frontmatter}

%\pacs{98.80.Cq, 98.70.Vc, 98.80.Es}
%\maketitle

%
\section{Introduction}\label{sec:intro}
Different processes in the early Universe may have generated a primordial gravitational wave background, such as, among others, quantum perturbations during inflation, cosmic strings, pre-big-bang theories, etc (for a complete review see \cite{Maggiore:2000gv} and references therein). Detecting this cosmological gravitational wave background (hereafter CGWB) provides a unique way to probe the primordial Universe and its evolution. \\
\indent The CGWB can be measured at low frequencies constraining a possible tensor-mode contribution to the large-scale temperature and polarization fluctuations in the cosmic microwave background radiation (hereafter CMB). The most recent constrain on the tensor-to-scalar ratio is the one published by the BICEP2/Keck joint analysis \cite{Array:2015xqh}, i.e. $r<0.07$ at 95\% CL \cite{Ade:2015xua,Ade:2015lrj}, which corresponds to $\Omega_{\rm gw} h^2 < 10^{-14}$ in the frequency range $10^{-17}-10^{-16} $ Hz.\\
At higher frequencies, in the range $10^{-9}-10^{-8} $ Hz, pulsars can be used as natural gravitational wave detectors, e.g. from the last European Pulsar Timing Array (EPTA) data release, Lentati et al. constrain the amplitude of GW up to  $\Omega_{\rm gw} h^2<1.1 \times 10^{-9}$ at 2.8 nHz \cite{Lentati:2015qwp}.
At smaller scales, interferometers such as LIGO \cite{Abbott:2007kv} and VIRGO \cite{Accadia:2011zzd} are also looking for gravitational wave signals. A recent bound, at $\sim 10^2$ Hz, is $\Omega_{\rm gw} h^2 <2.6 \times 10^{-6}$ from the cross correlation between LIGO and VIRGO detectors \cite{Aasi:2014zwg}. Most recently, during the review process of this paper, the LIGO-VIRGO collaboration detected a Gravitational Waves signal from a Binary Black Hole marger \cite{Abbott:2016blz}. The current best limit on stochastic background coming from Binary Black Holes is $\Omega_{\rm gw} (f=25 \, \text{Hz}) = 1.1 _{-0.9}^{+2.7} \times 10^{-9}$ at $90 \%$ CL \cite{TheLIGOScientific:2016wyq}.\\
\indent Moreover, at frequencies greater than $\sim 10^{-10}$ Hz, the stochastic background can be constrained through big-bang nucleosynthesis (BBN). In fact, at these frequencies, primordial gravitational waves contribute to the total radiation energy density, increasing the expansion rate of the Universe. In this scenario, the CGWB behaves as a free-streaming gas of massless particles. By measuring primordial abundances of light elements is possible to constrain the total number of relativistic degrees of freedom and, consequently, the gravitational waves energy density, for scales smaller than the comoving horizon at the end of the BBN \cite{Allen:1996vm}. \\
\indent Straightforwardly, it is possible to constrain the total radiation density through the CMB, reaching even smaller frequencies, $\sim 10^{-15}$ Hz, corresponding to the comoving horizon at the decoupling. In particular, if the CGWB energy density perturbations are adiabatic, the extra energy contribution due to gravitational waves is indistinguishable from the one due to relativistic neutrinos. Therefore, if we fix the relativistic degrees of freedom to its standard value, $N_{\text{eff}}=3.046$, and parametrize all the extra radiation as the effective number of gravitational waves degrees of freedom, $N_{\text{gw}}$, it is then easy to translate $N_{\text{gw}}$ into a CGWB energy density, as pointed out in \cite{Smith:2006nka}.\\
%As pointed out in \cite{Smith:2006nka}, it is easly possible to relate $N_{\text{gw}}$ to the CGWB energy density.\\%, considering that the contribution from a single massless neutrino species to a monochromatic CGWB is $\Omega_{\text{gw}} h^2=5.6\times10^{-6}$.\\
\indent In this paper we update previous constraints on the CGWB energy density, as those  presented in \cite{Sendra2012,Henrot-Versille:2014jua}, in light of the latest Planck data release and we present the bound reachable combining the future satellite missions COrE and Euclid. 

\section{Cosmological constraints on GW background}\label{sec:CMB}

In this section we discuss the datasets used in the analysis and the obtained results. We make use of the latest CMB Temperature and Polarization power spectra of the Planck survey \cite{Adam:2015rua,PlanckLikelihood} together with the Planck lensing likelihood \cite{Ade:2015zua}, the  Baryonic Acoustic Oscillations observations of \cite{Anderson} and the most recent primordial Deuterium abundance observation by Cooke et al. \cite{Cooke:2013cba}. Regarding CMB polarization data at large angular scales, we considered both the lowP likelihood based on 70~GHz data, described in \cite{PlanckLikelihood}, and the new SimLow likelihood based on 100 and 143~GHz HFI channels as described in \cite{Aghanim:2016yuo} by including an external prior on the optical depth $\tau=0.055 \pm 0.009$ at $68 \% $ CL.\\
\indent We explore the cosmological parameters space with the July 2015 version of the publicly available \texttt{cosmomc} package \cite{Lewis:2002ah}. We adopt the following parametrization for the $\Lambda$CDM model: the baryon and cold dark matter densities $\omega_b\equiv \Omega_{ b}h^2$ and $\omega_c\equiv \Omega_{ c}h^2$, the ratio of the sound horizon  to the angular diameter distance at decoupling $\theta_{MC}$, the re-ionization optical depth $\tau$, the scalar spectral index $n_S$, and the overall normalization of the spectrum $A_S$ at $k=0.05\, \text{Mpc}^{-1}$. Furthermore we assume adiabatic initial conditions and we impose spatial flatness. We fix the relativistic degrees of freedom, parametrized as $N_{\text{eff}}$, to its standard value of $3.046$ and, for each MCMC sample, we compute the primordial Helium abundance assuming standard BBN \cite{Iocco}, using a recent fitting formula based on results from the \texttt{PArthENoPE} BBN code \cite{parthenope}.\\
\indent As mentioned above, we parametrize the extra relativistic content due to the CGWB as $N_{\text{gw}}$, adding it to the total amount of massless neutrinos. To translate $N_{\text{gw}}$ to the CGWB energy density, we assume that the contribution from a single massless particle to a monochromatic CGWB is $ 5.6\times 10^{-6}$, therefore (following \cite{Maggiore:2000gv,Smith:2006nka} and what is done in \cite{Maggiore:1999vm}):
\begin{equation}\label{eq1}
\Omega_{\text{gw}} h^2 \equiv h^2 \int _0 ^{\infty} d(\ln f) \Omega_{\text{gw}}(f) = 5.6\times 10^{-6}N_{\text{gw}}
\end{equation}

\begin{figure}[H]
\centering
\includegraphics[scale=0.45]{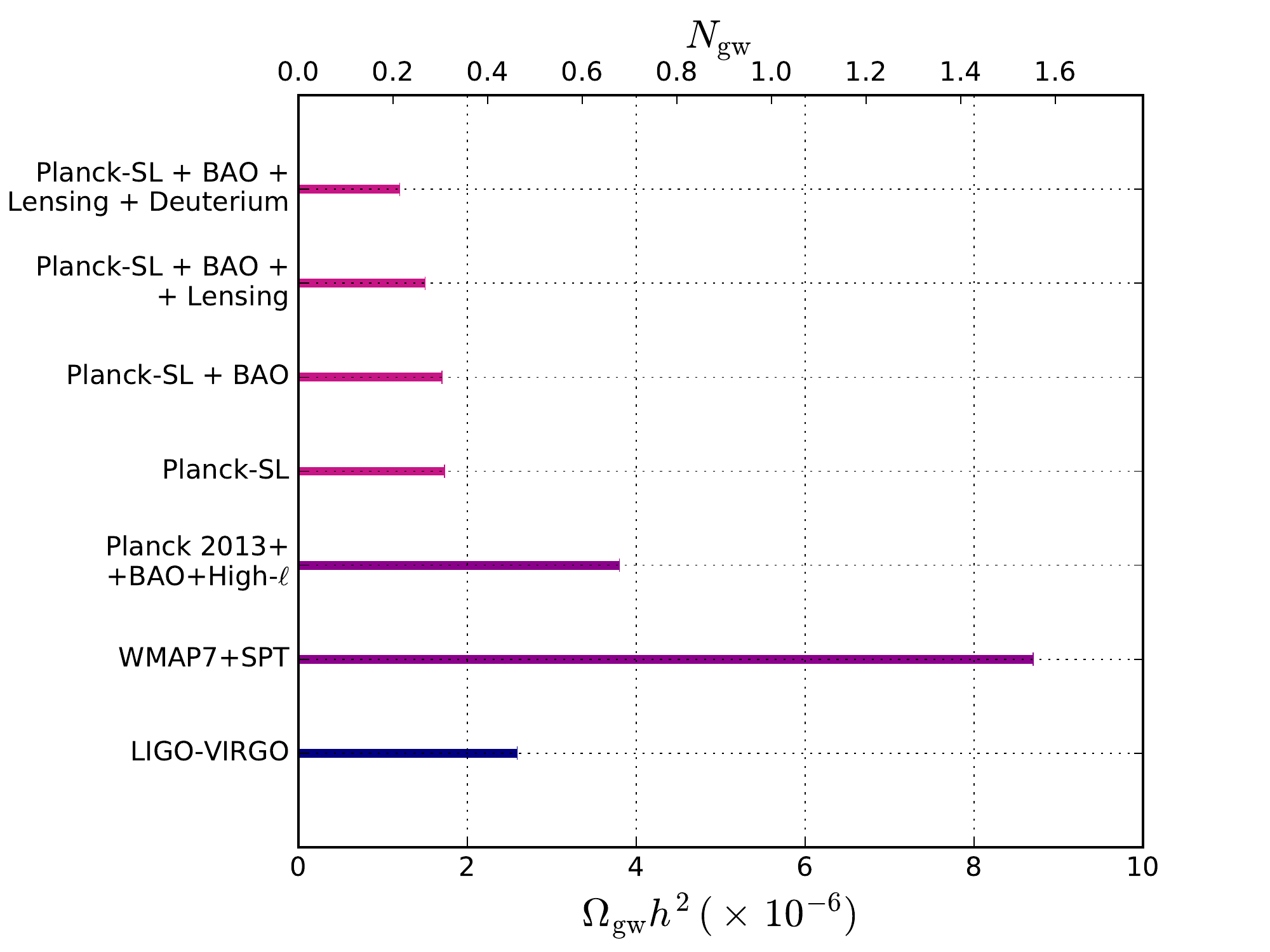}
\caption{\footnotesize{95\% CL upper limits for the CGWB energy density, $\Omega_{\text{gw}} h^2$ and for the effective number of gravitational waves degrees of freedom, $N_{\text{gw}}$ for the different analyzed datasets. With Planck-SL we refer to the Planck temperature and polarization Likelihood in combination with the new value of $\tau$ published in \cite{Aghanim:2016yuo}. We report here also the value quoted by the LIGO-VIRGO collaboration \cite{Aasi:2014zwg}, the previous cosmological constraint obtained combining WMAP7 \cite{WMAP7} and SPT\cite{SPT}}, and the Planck 2013 constraint \cite{Henrot-Versille:2014jua}.}
\label{fig:whisker}
\end{figure}

The effective number of gravitational waves degrees of freedom, acting as extra radiation, is completely degenerate with the extra relativistic degrees of freedom. Therefore, our bounds are model dependent, in fact, any process, that modifies the standard radiation content of the Universe, will affect the  $N_{\text{gw}}$ bounds. For example, the presence of extra relativistic particles at recombination (e.g. sterile neutrinos \cite{Abazajian:2012ys}) will tighten the bounds up. On the other hand, a reheating phase at $\MeV$ temperature can produce a value of $N_{\text{eff}}$ lower than the standard one \cite{Kawasaki:1999na}, relaxing the bounds on $N_{\text{gw}}$.\\
To translate the value of $\Omega_{\text{gw}} h^2$ in to a corresponding value of the tensor-to-scalar ratio $r$ is not straightforward since the shape of the tensor spectrum needs to be specified. For recent and very comprensive discussions on this specific topic see \cite{Meerburg:2015zua,Cabass:2015jwe}.\\
\indent We test different combinations of data, starting from Planck 2015 CMB data alone (both using lowP and SimLow likelihoods) and then adding BAO observations, Lensing data and the primordial Deuterium abundance measurement. We show in figure \ref{fig:whisker} the $95\%$ CL upper limits for the  CGWB energy density $\Omega_{\text{gw}} h^2$. 

\begin{table}[H]
\begin{center}
\scalebox{0.8}{
\begin{tabular}{l||c|c}
%\hline
\rule[-2mm]{0mm}{6mm}
Datasets & $\Omega_{\text{gw}} h^2$ & $N_{\text{gw}}$ \\
\hline
\hline
\rule[-2mm]{0mm}{6mm}
PlanckTTTEEE - lowP & $ < 2.1 \times 10^{-6} $  & $ < 0.37$\\
\hline
\rule[-2mm]{0mm}{6mm}
PlanckTTTEEE - lowP + BAO & $ <1.9  \times 10^{-6} $  & $ <0.34 $\\
\hline
\rule[-2mm]{0mm}{6mm}
PlanckTTTEEE - lowP + BAO + Lensing & $ < 1.6 \times 10^{-6} $  & $ < 0.29$\\
\hline
\rule[-2mm]{0mm}{6mm}
PlanckTTTEEE - lowP + BAO + Lensing + Deut. & $ < 1.2 \times 10^{-6} $  & $ < 0.22$\\
\hline
\hline
\rule[-2mm]{0mm}{6mm}
PlanckTTTEEE - SimLow & $ < 1.7 \times 10^{-6} $  & $ < 0.31$\\
\hline
\rule[-2mm]{0mm}{6mm}
PlanckTTTEEE - SimLow + BAO & $ < 1.7 \times 10^{-6} $ & $ <0.30 $ \\
\hline
\rule[-2mm]{0mm}{6mm}
PlanckTTTEEE - SimLow + BAO + Lensing & $ < 1.5 \times 10^{-6} $ & $ <0.27 $ \\
\hline
\rule[-2mm]{0mm}{6mm}
PlanckTTTEEE - SimLow + BAO + Lensing + Deut. & $ < 1.2 \times 10^{-6} $ & $ <0.22 $ \\
\hline
\hline
\rule[-2mm]{0mm}{6mm}
LIGO-VIRGO \cite{Aasi:2014zwg} &  $ <  2.6 \times 10^{-6} $  &   \\
\hline
\hline
\rule[-2mm]{0mm}{6mm}
COrE & $ < 0.50 \times 10^{-6} $ & $ <0.089 $ \\
\hline
\rule[-2mm]{0mm}{6mm}
COrE + Euclid & $ < 0.076 \times 10^{-6} $ & $ <0.013 $ \\
%\hline
\end{tabular}}
\caption{\footnotesize{95\% CL upper limits for the CGWB energy density $\Omega_{\text{gw}} h^2$ and the effective number of gravitational waves degrees of freedom ($N_{\text{gw}}$) for the considered cosmological datasets. We report also bounds from LIGO-VIRGO collaboration, in the frequency range $41.5 - 169.25$ Hz \cite{Aasi:2014zwg}. With lowP we refer to the likelihood published by Planck in 2015 \cite{PlanckLikelihood} while with SimLow we refer to the new value of optical depth published by Planck collaboration \cite{Aghanim:2016yuo}.}}
\label{tab:results}
\end{center}
\end{table}

In Table \ref{tab:results} we report the obtained upper limits on $\Omega_{\text{gw}} h^2$ and $N_{\text{gw}}$ for all the data combination that we consider. As expected using the new value of $\tau$ from Planck we reach more stringent constrains. On the other hand it is worth to notice that, including the Deuterium abundance measurements, this difference is practically negligible.\\
\indent The Planck-SimLow alone constraint is 50\% better than the upper bounds estimated by the LIGO-VIRGO collaboration in 2014 \cite{Aasi:2014zwg}. Combining CMB power spectra with the Lensing likelihood and the BAO data we obtain a slightly more stringent constraint, $\Omega_{\text{gw}} h^2< 1.5\times 10^{-6}$; finally, assuming standard BBN \cite{Iocco,parthenope}, we add also the primordial Deuterium abundance measurement reaching $\Omega_{\text{gw}} h^2< 1.2\times 10^{-6}$  at 95\% CL,  improving the precedent Planck constraint by a factor of about $3$ \cite{Henrot-Versille:2014jua}, the pre-Planck cosmological constraint \cite{Sendra2012} (based on WMAP \cite{WMAP7}  and South Pole Telescope \cite{SPT} results) by a factor of $6$ and the current interferometer measurements by about $2$.
\indent We also verified the stability of our results with respect to assumptions on massive neutrinos. By opening $\sum m_{\nu}$ as extra parameter, we found, as expected, an overall relaxation on $N_{\text{gw}}$ bounds. Nevertheless for all the data combinations we have considered above, the upper limits increase less than $10 \%$, not changing our conclusions.\\
\indent It is also interesting to forecast the future sensitivity on $\Omega_{\text{gw}} h^2$  achievable with future satellite missions such as COrE \cite{COrE,COrE++} and Euclid \cite{Euclid}. To this end we simulate mock data for the COrE mission using the 5 channels in the range 100-220 GHz, following the approach described in \cite{Lewis:2005tp}, assuming perfect foreground removal and ignoring correlations between multipoles. Analyzing  this dataset we find that the COrE mission will be able to constrain the CGWB energy density to $\Omega_{\rm gw} h^2 <5.0\times 10^{-7} $ at 95\% CL. For the Euclid mission we use the fisher matrix approach as described in \cite{Martinelli,Amendola}; we then combine the inverse covariance matrix produced by \texttt{cosmomc} for COrE with the Euclid fisher matrix obtaining $\Omega_{\rm gw} h^2 <7.6\times 10^{-8} $ at 95\% CL.

\section{Conclusions}\label{sec:conclusions}

We have used the latest Planck data to constrain a possible cosmological gravitational wave background at frequencies greater than $10^{-15}$ Hz. Our tighter constraint is $\Omega_{\rm gw} h^2 <1.2\times 10^{-6} $ at 95\% CL, obtained combining CMB with BAO, Lensing and primordial Deuterium observations. This result improves previous cosmological bounds by a factor $3$ and the recent LIGO-VIRGO direct measurements by $2$.\\
\indent We also show that with the next generation cosmological satellite missions (COrE and Euclid) would be possible to shrink the bounds by more than one order of magnitude with respect to current limits.
The constraints presented here are probably not significant for slow roll inflation that produces essentially scale invariant spectra.
Those models are already strongly constrained by current large scale bounds on primordial CMB polarization B-modes.
However, phase transitions, pseudo scalar inflation  or other exotic mechanisms that produce a CGWB at higher frequencies (see for example \cite{kadota} and \cite{pseudoscalar}) can be constrained by the bounds presented here.
$ $\\

%\section{Acknowledgments}
\noindent {\bf Acknowledgements}

We are grateful to G. Cabass and M. Lattanzi for useful discussions and suggestions. We acknowledge the use of computing facilities at NERSC (USA). We acknowledge partial financial support by the research grant Theoretical Astroparticle Physics number 2012CPPYP7 under the program PRIN 2012 funded by MIUR and by TASP, iniziativa specifica INFN.

%%%%%%%%%%%%%%%%%%%%%%%%%%%%%%%%%%%%%%%%%%%%%%%
%%%%%%%%%%%%%%%%%%%%%%%%%%%%%%%%%%%%%%%%%%%%%%%

\end{document}